\documentclass[12pt,a4paper,final]{iopart}

\usepackage{iopams}  
\usepackage{graphicx}
\usepackage[breaklinks=true,colorlinks=true,linkcolor=blue,urlcolor=blue,citecolor=blue]{hyperref}

\begin{document}

\title[All-optical production of a superfluid Bose-Fermi mixture of $^6$Li and $^7$Li]{All-optical production of a superfluid Bose-Fermi mixture of $^6$Li and $^7$Li}

\author{Takuya Ikemachi$^1$, Aki Ito$^2$, Yukihito Aratake$^1$, Yiping Chen$^1$, Masato Koashi$^{2,3,4}$, Makoto Kuwata-Gonokami$^{1,3,4}$, and Munekazu Horikoshi$^{3,4}$}
\address{$^1$Department of Physics, Graduate School of Science, The University of Tokyo, 7-3-1 Hongo, Bunkyo-ku, Tokyo 113-0033, Japan.}
\address{$^2$Department of Applied Physics, Graduate School of Engineering, The University of Tokyo, 7-3-1 Hongo, Bunkyo-ku, Tokyo 113-8656, Japan.}
\address{$^3$Institute for Photon Science and Technology, Graduate School of Science, The University of Tokyo, 7-3-1 Hongo, Bunkyo-ku, Tokyo 113-8656, Japan.}
\address{$^4$Photon Science Center, Graduate School of Engineering, The University of Tokyo, 2-11-16 Yayoi, Bunkyo-ku, Tokyo 113-8656, Japan.}

\begin{abstract}
We report the first all-optical production of a superfluid Bose-Fermi mixture with two spin states of $^6$Li (fermion) and one spin state of $^7$Li (boson) under the resonant magnetic field of the $s$-wave Feshbach resonance of the fermions.
Fermions are cooled efficiently by evaporative cooling and they serve as coolant for bosons.
As a result, a superfluid mixture can be achieved by using a simple experimental apparatus and procedures, as in the case of the all-optical production of a single Bose-Einstein condensate (BEC).
We show that the all-optical method enables us to realize variety of ultracold Bose-Fermi mixtures.

\end{abstract}


\section{Introduction}

Quantum many-body systems of strongly interacting fermions are ubiquitous in condensed matter physics, nuclear physics, astrophysics, and atomic physics. Tunable interactions between particles through the use of a Feshbach resonance have enabled us to simulate such Fermi systems in cold atom experiments. To date, many-body physics for two-component fermions have been studied in the absence of impurities \cite{ref1,ref2}. However, realistic physical systems are not such a pure Fermi system. For example, phonons, holes, and impurities give rise to unique features in condensed matter at low temperature.

A Bose-Fermi mixture is an example of a system that allows us to investigate fermions interacting with other particles. Historically, Bose-Fermi mixtures have been studied for the purpose of cooling fermions via sympathetic cooling \cite{ref3,ref4}. However, recent breakthroughs in realizing superfluid Bose-Fermi mixtures of $^6$Li and $^7$Li \cite{ref5} have completely changed the objective of studying such mixtures. Superfluid Bose-Fermi mixtures have the potential to reveal novel quantum phenomena such as the phase diagram and stability of structures \cite{ref6,ref7,ref8,ref9}, changes in the dispersion relation of Bose-Einstein condensates (BECs) \cite{ref10}, the lifetime of quasiparticles, the damping rates of elementary excitations \cite{ref10,ref11,ref12}, counterflows of two superfluids and their critical velocities \cite{ref12}, and quantum phase transitions in Bose-Fermi mixtures \cite{ref13,ref14}.

In this paper, we report the first all-optical production of a superfluid Bose-Fermi mixture of $^6$Li and $^7$Li. Previously, all-optical methods have been developed to realize a BEC \cite{ref15} or a degenerate Fermi gas efficiently using a simple experimental apparatus \cite{ref16}. These methods are also compatible with the use of Feshbach resonance. However, until now, a superfluid Bose-Fermi mixture had not yet been obtained using all-optical methods.

\section{Experimental setup}

\begin{figure}[tb!]
 \centering
 \includegraphics{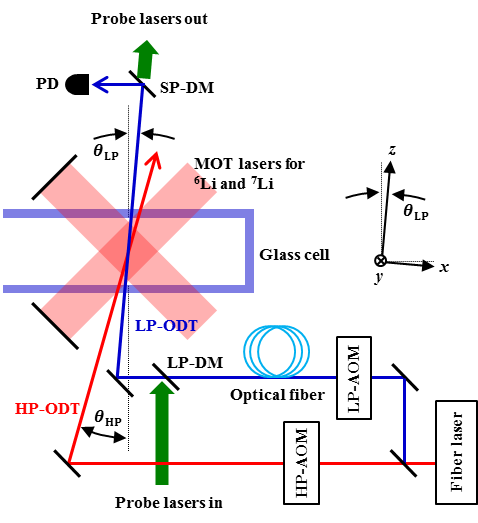}
 \caption{\label{fig1}
Top view of the experimental setup and schematic drawing of the optical setup. The $z$ axis is defined as the axial direction of the LP-ODT. The axis of the magnetic field is the $y$ direction. PD, LP-DM, and SP-DM are a photodetector, longpass dichroic mirror, and shortpass dichroic mirror, respectively. The lenses and wave plates used in the setup are not shown.
}
\end{figure}

A part of the experimental setup is shown in Fig.~\ref{fig1}. $^6$Li and $^7$Li are pre-cooled above a lithium oven using a two-dimensional MOT (2D-MOT), as shown in Ref.~\cite{ref17}. They are loaded into a simultaneous three-dimensional magneto-optical trap (3D-MOT) \cite{ref18} in a glass cell through a differential pumping tube. Cooling and repumping lasers for the simultaneous MOT have $1/e^2$ radii of 5~mm, and their laser power are $P_{\rm cool}^6 = 42$~mW, $P_{\rm rep}^6 = 38$~mW, $P_{\rm cool}^7 = 34$~mW, and $P_{\rm rep}^7 = 22$~mW. As suggested in Ref.~\cite{ref18}, we use the D1 line only for $^6$Li repumping and the D2 lines for other transitions in order not to heat $^7$Li. The quadrupole magnetic field for the MOT is set to 18~Gauss/cm in the strong direction (the $y$ direction in the figure).

The optical setup for the optical dipole trap (ODT) is illustrated in Fig.~\ref{fig1}. We use a double ODT system that can produce a deep ODT and a stable ODT \cite{ref19}. The light source of this double ODT system is a multimode ytterbium fiber laser with a maximum output power of 200~W, a center wavelength of 1070~nm, and a line width of 3~nm. The output is divided into two optical paths. One is used for the deep ODT produced with high laser power (HP-ODT). The power is controlled using an acousto-optic modulator (AOM) with a large active aperture of 2.5$\times$2.5~mm$^2$ manufactured in crystal quartz with a small absorptance of 0.4\% (HP-AOM). The other path is used for the stable ODT produced by low laser power (LP-ODT). The power is controlled using an AOM with an active aperture of 1$\times$1~mm$^2$ manufactured in TeO$_2$ (LP-AOM). The diffracted laser is coupled into an optical fiber for spatial filtering and pointing stability. Both laser beams are focused to the MOT region in the glass cell with beam waists of $w_{\rm HP}$ = 37~$\mu$m and $w_{\rm LP}$ = 45~$\mu$m. They are orthogonally polarized relative to each other to avoid interference at their intersection. The finite incident angles of $\theta_{\rm HP}$ = 16$^{\circ}$ and $\theta_{\rm LP}$ = 5$^{\circ}$ are chosen in order to prevent back reflections inside the glass cell from returning to the trapping region. The glass cell is made from anhydrous quartz to reduce thermal lens effects caused by the high-power laser ODT. The laser power of the LP-ODT is stabilized by monitoring the power after the glass cell and feeding the error signal back to the LP-AOM. Typically, a laser power of 100~W (2~W) can be delivered inside the glass cell used for the HP-ODT (LP-ODT) under an output of 160~W from the fiber laser. The maximum depth of these ODTs are 2.8~mK and 38~$\mu$K, respectively. The measured 1/e lifetime of the LP-ODT is 60~s.

A bias magnetic field for the Feshbach resonance of $^6$Li is produced by the same pair of coils used for the MOT. They produce a magnetic curvature in the $z$ direction, corresponding to 0.24$\sqrt{B}$~Hz for $^6$Li and $\eta \times$0.24$\sqrt{B}$~Hz for $^7$Li, where $B$ is the bias magnetic field produced by the coils in units of Gauss. $\eta$=$\sqrt{m_6/m_7}\sim$0.93 is a factor for the mass difference. We omit magnetic curvature in the $x$ and $y$ directions since atomic confinement by the ODTs is dominant in these two directions. The total trapping potential is given by a combination of the ODTs and this magnetic confinement.

The probe laser beams for $^6$Li and $^7$Li are aligned on the path of the LP-ODT and applied along the $z$ direction for the measurement of the momentum distributions. Additionally, we apply probe beams along the $y$ direction to measure the {\it in-situ} density distributions of atoms \cite{ref20}.

\section{Experimental procedure}

\begin{figure}[tb!]
 \centering
 \includegraphics{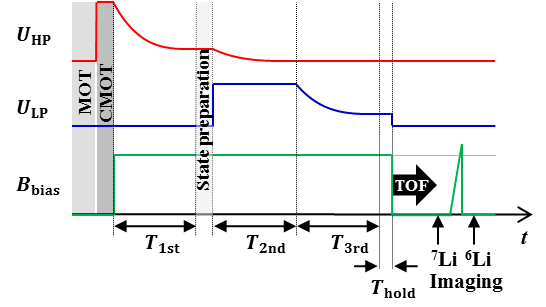}
 \caption{\label{fig2}
Experimental procedure. From top to bottom, the solid lines show the time variations of the trap depth of the HP-ODT, the LP-ODT, and the bias magnetic field, respectively.
}
\end{figure}

The experimental procedure is shown in Fig.~\ref{fig2}. We collected 2$\times$10$^8$ $^6$Li atoms and 2$\times$10$^8$ $^7$Li in the simultaneous MOT within 30~s. We loaded them into the HP-ODT by further cooling and compressing in a compressed-MOT (CMOT) for 45~ms. During the CMOT process, we gradually increased the MOT field up to 90~Gauss/cm, decreased the power of the MOT lasers, and changed the detuning of the MOT lasers in the presence of an HP-ODT with a depth of 2.8~mK. Approximately 1$\times$10$^7$ $^6$Li atoms and 3$\times$10$^6$ $^7$Li atoms are loaded into the HP-ODT in this loading process. This difference in loading efficiency between $^6$Li and $^7$Li is caused by a small position displacement of $^7$Li MOT from $^6$Li MOT.

Immediately after the CMOT, a bias magnetic field of 832.18~Gauss is turned on for subsequent evaporative cooling. At the beginning of evaporative cooling, $^6$Li atoms are populated in the two magnetic sublevels of the lowest hyperfine state $F=1/2$, and $^7$Li atoms are populated in the three magnetic sublevels of the $F = 1$ state. At this magnetic field, fermions have a diverging scattering length of $|a_{\rm ff}|=\infty$ for collisions between the two states \cite{ref21}, and bosons have $a_{\rm bb}^{+1}=-60a_0$ for $m_F=+1$ \cite{ref22,ref23}, $a_{\rm bb}^0=70a_0$ for $m_F=0$ \cite{ref5,ref24}. The scattering length of $a_{\rm bb}^{-1}$ for $m_F=-1$ is unknown.
In this experiment, we do not consider the scattering length between different internal states of the bosons, because we choose one of these internal states in the process of evaporative cooling. The scattering lengths between bosons and fermions have almost the same values of $a_{\rm bf}\sim 40a_0$ for all the combinations of internal states \cite{ref3,ref5}. Since $|a_{\rm ff} |$ $\gg$ $|a_{\rm bb}|$, $|a_{\rm bf}|$, fermions can be cooled more efficiently than bosons by  evaporative cooling. Thus, bosons can be cooled by thermal contact with cooled fermions, namely, by sympathetic cooling. This process is the reverse of the sympathetic cooling of fermions by bosons \cite{ref3,ref4}.

Evaporative cooling was performed in the double ODT system. First, the trap depth of the HP-ODT was lowered from 2.8~mK to 560~$\mu$K for a period of $T_{\rm 1st}$ by decreasing the output laser power of the fiber laser from 160~W to 33~W, which changes the power inside the glass cell from 100~W to 20~W. We then applied a radio-frequency magnetic pulse for 300~ms to $^6$Li with a resonant frequency between the two magnetic sublevels of $F = 1/2$ in order to populate them equally and maximize the efficiency of evaporative cooling. The $m_F = +1$ state of $^7$Li disappeared during evaporative cooling owing to inelastic collisions among the bosons at this magnetic field, leaving the stable internal states $m_F = 0$ and $m_F = -1$. We prepared one of these internal states by applying a blower laser pulse to $^7$Li atoms for 50~$\mu$s with a frequency tuned to $m_F = 0$ or $m_F = -1$ to remove them selectively from the trap. After the state preparation process, we turned on the LP-ODT with a depth of 38~$\mu$K (2~W) and produced a crossed-ODT at the intersection of the HP-ODT and LP-ODT and gradually turned off the HP-ODT for a period of $T_{\rm 2nd}$. During this step, atoms were cooled and transferred into the stable LP-ODT. They were finally cooled by lowering the trap depth down to 850~nK (45~mW) for a period of $T_{\rm 3rd}$ and held for a period of $T_{\rm hold}$ until they reached thermal equilibrium. The final trap has trapping frequencies of $\nu_r = 240$~Hz and $\nu_z = 7.1$~Hz for $^6$Li in the radial and axial directions, respectively. The frequencies for $^7$Li were obtained by multiplying those for $^6$Li by a factor of $\eta$.

The final state of the Bose-Fermi mixture was examined by measuring the momentum distributions of the paired fermions and the bosons. We turned off the LP-ODT and the bias magnetic field simultaneously (within 10~$\mu$s of each other). After the $^7$Li atoms expanded ballistically during 8~ms of the time of flight (TOF), the probe laser beam was applied for absorption imaging. In contrast, the paired fermions of $^6$Li atoms were converted into molecules when the magnetic field was turned off \cite{ref25,ref26}, and those molecules expanded with the original center-of-mass momentum distribution of the paired fermions. Although unpaired fermions also expanded as atoms or as randomly associated molecules, they exhibited a thermal distribution. Therefore, macroscopic occupation at the zero momentum state in the molecules suggested the existence of condensate paired fermions in the Fermi system before the magnetic field was turned off \cite{ref27}. Instead of imaging the molecules directly, they were dissociated into atoms by applying a magnetic pulse with a larger peak than the resonant magnetic field strength of the Feshbach resonance. In this way, the momentum distribution of the paired fermions and unpaired atoms were imaged after the total TOF time of 11~ms.

\section{Experimental results}

\begin{figure}[tb!]
 \centering
 \includegraphics{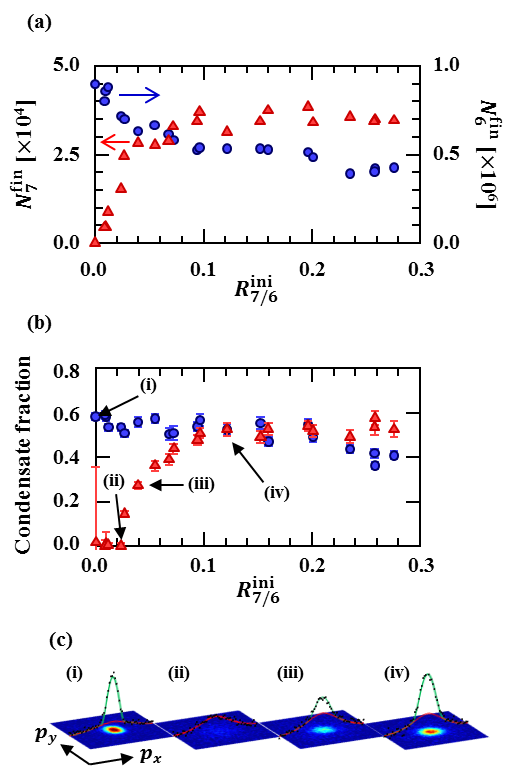}
 \caption{\label{fig3}
Bose-Fermi mixture of $^6$Li in the two magnetic sublevels of $F=1/2$ and $^7$Li in the $m_F=0$ state of $F=1$. The scattering lengths are $|a_{\rm ff}|=\infty$, $a_{\rm bb}^0=70a_0$, and $a_{\rm bf}\sim40a_0$ at 832.18~Gauss. The experimental conditions are $T_{\rm 1st}=5$~s, $T_{\rm 2nd}=4$~s, $T_{\rm 3rd}=20$~s, $T_{\rm hold}=1$~s, and $N_6^{\rm ini}=1\times 10^7$. (a) Final number of $^7$Li (triangles) and $^6$Li (circles) as a function of a mixing rate $R_{7/6}^{\rm ini}$. (b) Condensate fraction of bosons (triangles) and paired fermions (circles) as a function of the mixing rate $R_{7/6}^{\rm ini}$. (c) Absorption images of the momentum distribution of paired fermions (i) and bosons (ii-iv). The indexes correspond to those indicated by the arrows in (b). The black dots show the momentum distributions integrated along the $y$ direction and the solid lines show the results of bimodal fitting.
}
\end{figure}

First, we show the experimental results of evaporative cooling for a mixture of fermions and bosons in the $m_F = 0$ state with $a_{\rm bb}^0=70a_0$. To maximize the final phase space density of $^7$Li, we optimized the set $T_{\rm 1st}$, $T_{\rm 2nd}$, $T_{\rm 3rd}$, and $T_{\rm hold}$. They were found to be 5~s, 4~s, 20~s, and 1~s, respectively. Fig.~\ref{fig3}a shows the final number of atoms after evaporative cooling as a function of the initial mixing ratio given by $R_{7/6}^{\rm ini}=N_7^{\rm ini}/N_6^{\rm ini}$, where $N_6^{\rm ini}$ and $N_7^{\rm ini}$ are the initial numbers of $^6$Li and $^7$Li at the beginning of evaporative cooling, respectively. Here, $N_6^{\rm ini}$ was fixed to $1\times 10^7$, and only $N_7^{\rm ini}$ was controlled by the power of the $^7$Li cooling laser in the MOT. Fig.~\ref{fig3}b shows the condensate fraction (CF) of the paired fermions and bosons, evaluated by fitting a bimodal distribution function to their momentum distributions (Fig.~\ref{fig3}c). When the $^7$Li atoms were not mixed ($R_{7/6}^{\rm ini}=0$), approximately $N_6^{\rm fin}=9\times 10^5$ $^6$Li atoms reached the superfluid state and the observed CF was approximately 0.6. This value is close to the maximum CF of the paired fermions at the unitarity limit \cite{ref28}. We can see that bosons can enter the BEC phase at $R_{7/6}^{\rm ini} > 0.02$. The fact that both $^6$Li and $^7$Li show finite CF at $R_{7/6}^{\rm ini} > 0.02$ provides direct evidence for the presence of a superfluid Bose-Fermi mixture. At $R_{7/6}^{\rm ini} = 0.12$, the superfluid mixture consisted of $3\times 10^4$ $^6$Li and $5\times 10^5$ $^7$Li.

\begin{figure}[tb!]
 \centering
 \includegraphics{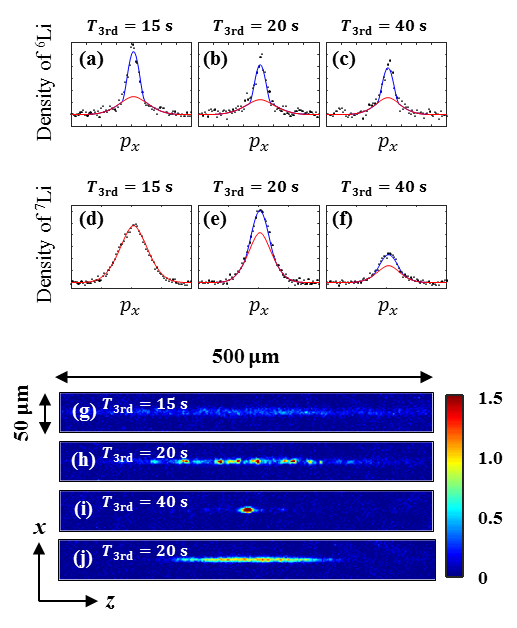}
 \caption{\label{fig4}
Bose-Fermi mixture of $^6$Li in the two spin states of $F=1/2$ and $^7$Li in the $m_F=-1$ state of $F=1$. The scattering lengths are $|a_{\rm ff}|=\infty$, $a_{\rm bb}^{-1}=-3.2a_0$, and $a_{\rm bf}\sim40a_0$ at 832.18~Gauss. The experimental conditions are $T_{\rm 1st}=5$~s, $T_{\rm 2nd}=4$~s, $T_{\rm hold}=1$~s, $N_6^{\rm ini}=1\times 10^7$, and $R_{7/6}^{\rm ini}\sim 0.3$. Momentum distributions of (a-c) paired fermions and (d-f) bosons at different $T_{\rm 3rd}$. The black dots show the momentum distributions integrated along the $y$ direction and the solid lines show the results of bimodal fitting. (g-i) {\it In-situ} density distributions of the bosons in $m_F=-1$ at different $T_{\rm 3rd}$. (j) {\it In-situ} density distribution of the bosons in $m_F=0$ at $T_{\rm 3rd}=20$~s.
}
\end{figure}

Next, we show the experimental results of evaporative cooling for a mixture of fermions and bosons in the $m_F = -1$ state with an unknown scattering length $a_{\rm bb}^{-1}$. We investigated the final state by changing $T_{\rm 3rd}$ with fixed parameters of $T_{\rm 1st}=5$~s, $T_{\rm 2nd}=4$~s, $T_{\rm hold}=1$~s, and $R_{7/6}^{\rm ini}\sim0.3$. The result is shown in Fig.~\ref{fig4}. When we chose $T_{\rm 3rd}=15$~s, fermions exhibited a bimodal distribution but bosons exhibited a thermal distribution (Fig.~\ref{fig4}a,d,g). At $T_{\rm 3rd}=20$~s, which is the optimum condition for $m_F=0$, boson and fermions both exhibited bimodal distributions in their momentum distributions (Fig.~\ref{fig4}b,e). This is evidence for a superfluid Bose-Fermi mixture. However, the {\it in-situ} density distribution of bosons showed fragmented BECs (Fig.~\ref{fig4}h), while the case of $m_F=0$ showed a single BEC (Fig.~\ref{fig4}j). When we prolonged $T_{\rm 3rd}$ up to 40~s, the bosons in $m_F=-1$ showed a single BEC (Fig.~\ref{fig4}i). We can see that the axial size of the BEC is much smaller than that of the repulsively interacting BEC of $m_F=0$ ($a_{\rm bb}^0=70a_0$).
We evaluated the scattering length to be $a_{\rm bb}^{-1}=-3.2(3)a_0$ at $B = 832.18$~Gauss from the number of atoms in the BEC: $N_{\rm atom}=7.1\times 10^3$, the axial size in the $z$ direction: $w_{\rm BEC}=6.1(3)$~$\mu$m, and the trapping frequencies according to \cite{ref29}.
The negative sign of the scattering length is consistent with the inequality of $w_{\rm BEC}<l_z=\sqrt{\hbar/m_7\omega_z}=14.8$~$\mu$m, where $l_z$ is the harmonic oscillator length.
Therefore, this is the first realization of a mixture of Fermi superfluid and an attractive BEC.
The observed fragmented BECs can be explained as the spontaneous creation of Kibble-Zurek solitons, which was studied in Ref.~\cite{ref30}.

When we prepared only $^7$Li without $^6$Li at the beginning of evaporative cooling, we could not increase the phase space density of $^7$Li even for a longer evaporation time. This means that the elastic scattering rate of $^7$Li determined by the initial phase space density and the scattering length is not sufficient for evaporative cooling. Thus, $^7$Li atoms in the Bose-Fermi mixture are mainly cooled by sympathetic cooling with fermions. Fig.~\ref{fig3}a also shows evidence for sympathetic cooling. The figure shows that $N_6^{\rm fin}$ decreases as $R_{7/6}^{\rm ini}$ increases. This could be because more $^6$Li atoms need to be evaporated to cool more $^7$Li atoms by sympathetic cooling.

\section{Summary}

We demonstrated an experimental technique to produce a superfluid Bose-Fermi mixture of $^6$Li and $^7$Li using an all-optical method. During evaporative cooling, fermions were cooled efficiently by evaporative cooling with an enhanced scattering rate, while bosons were successfully cooled by sympathetic cooling with the aid of fermions. As a result, such a fascinating mixture can be produced using a simple experimental apparatus, as in the case of all-optical production of a single BEC. We demonstrated various Bose-Fermi mixtures by controlling the number, spin state, and cooling speed of bosons. Such high controllability paves the way for future studies on many-body physics of superfluid fermions with impurities.

\section*{Acknowledgments}

The authors would like to thank J. Metz, K. Togashi, and T. Otsu for supporting the initial stage of this work. This work was supported by a Grant-in-Aid for Scientific Research on Innovative Areas (Grant No. 24105006) and a Grant-in-Aid for Young Scientists (A) (Grant No. 23684033).

\end{document}